\documentclass[a4paper,12pt]{article}
\begin{document}
\begin{center}
{\noindent \large {\bf A new analytical solving for electric \\ polarizabilities 
of hydrogen-like atoms}} \\[.3in]
{\bf {\normalsize V. F. Kharchenko}} 
 \\[.05in]
\noindent {\footnotesize {\it Bogolyubov Institute for Theoretical Physics, \\ 
National Academy of Sciences of Ukraine, UA - 03143, Kyiv, Ukraine \\ 
vkharchenko@bitp.kiev.ua}}  \\[.1in]
\end{center}
{\small The direct transition-matrix approach to the description of the electric polarization 
of the quantum bound system of particles is used to determine the 
electric multipole polarizabilities of the hydrogen-like atoms. It is 
shown that in the case of the bound system formed by the Coulomb 
interaction the corresponding inhomogeneous integral equation determining 
an off-shell scattering function, which consistently describes 
virtual multiple scattering, can be solved exactly analytically for all 
electric multipole polarizabilities. Our method allows to reproduce the known
Dalgarno-Lewis formula for electric multipole polarizabilities of the 
hydrogen atom in the ground state and can also be applied to determine 
the polarizability of the atom in excited bound states. \\ [.05in]
[.05in]
} 
\\ [.2in] 

\noindent {\bf 1. Introduction} \\ 

The phenomenon of polarization (deformation) under the action of an 
external electromagnetic field is inherent to all composite quantum 
systems containing one or more charged components (to molecules, atoms, 
nuclei, nucleons and other complexes) and is indisputably an universal 
property of the system. 

Theoretical studies of influence of the electric and magnetic fields 
on structure of spectra of the simplest atoms were in the center of 
attention while becoming of modern quantum mechanics [1,2]. The Stark 
effect of the second order for hydrogen-like atoms was first 
calculated by Epstein [3], Wentzel [4] and Waller [5] based on the 
separation of the Schr\"{o}dinger equation in parabolic coordinates 
and by using the perturbation theory.

Later on, the electric dipole polarizability of the simplest atomic 
systems were calculated directly using the traditional method by the 
formula for the energy shift in the second order of the 
Rayleigh-Schr\"{o}dinger perturbation theory [2,6,7] in the applied 
electric field. As an example, employing this method, the contributions 
of the excited virtual states to the electric dipole polarizability 
of the hydrogen atom were investigated in Ref. 8. However, in general 
case this method is too cumbersome and nonpracticable because of the 
necessity of taking into account of virtual excited states, both 
discrete and from the continuum, that are coupled with the ground 
state by the perturbated interaction, especially for the systems with 
three and more particles.

The modification of the Rayleigh-Schr\"{o}dinger perturbation theory 
proposed by Dalgarno and Lewis [9] (see also [10-12]) permits to avoid 
difficulties associated with allowance for all possible intermediate 
bound and continuum states by the prior determination of the 
first-order correction to the wave function of the unperturbated 
state which satisfies an inhomogeneous differential equation. 
In the case of the hydrogen atom Dalgarno and Lewis managed to solve 
analytically the differential equation and first derive a general 
formula for all electric multipole polarizabilities of the atom in 
the ground bound state. With the use of the O(4)-symmetry of the 
energy operator of the hydrogen-like atom [13] and Dalgarno-Lewis 
perturbative technique [9] purely algebraic approach to the calculation 
of the second-order Stark effect for the atom has been developed in Ref. [14].

The object of this paper is to elaborate a new method of the analytical 
solution of the electric multipole polarizabilities of the hydrogen-like 
atom on the basis of the transition matrix approach, which has been  
applied by us earlier in nuclear physics to calculate the electric 
dipole polarizabilities of systems with the $S$-wave finite-range 
interaction --- the deuteron [15-18] and two-cluster models of the 
triton and lambda hypertriton [19,20]. In the preceding paper [21] 
the t-matrix formalism was firstly applied to calculate numerically 
the dipole, quadrupole and octupole polarizabilities of the hydrogen atom 
using the representation of the Coulomb t-matrix with explicitly 
removed singularities [22].

In this paper we demonstrate that the integral equations, 
which appear in the framework of the t-matrix formalism, 
permit in the case of the Coulomb interaction exact analytical 
solution for all the electric multipole polarizabilities of the 
hydrogen-like atom. Section 2 is devoted to description of the 
t-matrix formalism of the polarization interaction of a two-partice 
bound complex placed in a external electric field of a charged particle. 
In Section 3 the proposed approach is applied for derivation of 
the main integral equation that determines the 
electric multipole polarizabilities of the hydrogen-like atom. 
Strictly analytical solving of the obtained integral equation 
is performed in Section 4 in the case of the ground bound state.  
In Section 5 the general formula for the electric multipole 
polarizabilities of the hydrogen-like atom in the ground state is 
derived and discussed. Section 6 is concerned with the analytical 
derivation of the electric dipole polarizability of the hydrogen-like 
atom in the excited $2S$-state. Conclusions and outlook are presented 
in Section 7. 
\\ [.2in]
\noindent {\bf 2. Transition matrix description of polarization interaction \\
for a two-particle bound complex} \\ 

Let us consider scattering of a two-particle complex formed from 
charged particles 1 and 2 in the ground bound state by an external 
electric field of a charged particle 0. The total Hamiltonian of 
the three-particle system is 
\begin{equation}
H = H^0 + V \;,
\end{equation}
where  $H^0 = h^0_{12} + h^0_0$ is the kinetic-energy operator of 
the system ($h^0_{12}$ is the operator of the kinetic energy of 
relative motion of the particles 1 and 2 and $h^0_0$ is
that of the particle 0 and the center of mass of the particles 
1 and 2) and $V = v_{12} +  v_{01} +  v_{02}$ is the potential-energy 
operator of the system, $v_{01}$ and  $v_{02}$ are the potentials of 
the pair (Coulomb) interaction between the particle 0 and 
the particles 1 and 2 of the complex,
\begin{equation}
  v_{0i} = \frac{e_0 e_i}{r_{0i}}\;,
\end{equation}
$e_1$ and $e_2$ are the charges of the constituents of the complex, 
$e_0$ is the charge of the particle 0 that is the source of the 
electric field, $r_{0i}$ is the relative distance between the 
particles 0 and $i$,  $v_{12}$ is the interaction potential that 
supports the formation of the bound complex.

Representing the potential $V$ as the sum of the 
"internal" interaction potential, $v_{12}\;$, and "external" one, $v_0\;$, 
\begin{equation}
 V = v_{12} + v_{0}\;,\qquad  v_{0} = v_{01} + v_{02}\;, 
\end{equation}
and applying the well-known Francis-Watson [23] and Feshbach [24, 25] 
technique (see also [26]) with the use of the projection operators 
$ P \equiv \mid \psi \rangle \langle \psi \mid $ and $ Q = 1 - P $, where
$\psi$ is the wave function of the two-particle complex in the 
bound state with the binding energy $b$ (normalized to 1) 
that satisfies the Schr\"{o}dinger equation 
\begin{equation}
(h^0_{12} + v_{12} + b)\psi = 0  ,
\end{equation}
we wright the effective interaction potential between the particle 0 and the 
complex in the form 
\begin{equation}
v_{eff} = \langle \psi \mid R(E) \mid \psi \rangle  .
\end{equation}
Three-particle operator $R(E)$ in (5) satisfies to the integral equation 
of the Lippmann-Schwinger type
\begin{equation}
 R(E) = v_0 + v_0 G^Q_{12}(E) R(E)\;, 
\end{equation}
in which the potential-energy operator is the "external" 
interaction potential describing the interaction between the 
constituent particles of the complex and the particle 0 (3), 
and the role of the propagator fulfils the "truncated" Green's operator 
that contains the "internal" interaction potential $v_{12}$,  
\begin{equation}
G^Q_{12}(E) \equiv Q G_{12}(E)\;, \qquad  G_{12}(E)= (E-H^0-v_{12})^{-1}\;,
\end{equation}
where $E=\varepsilon - b$ is the total energy of the system, 
$\varepsilon$ is the energy of the relative motion of the 
center of mass of the complex and the particle 0. The operator 
$G_{12}(E)$ is determined by the integral equation 
\begin{equation}
G_{12}(E)= G^0(E) + G^0(E) v_{12} G_{12}(E)\;.
\end{equation}
where  $G^0(E)$ is the free three-body Green's operator,
$$
 G^0(E) = (E - H^0)^{-1}.
$$
Introducing  the transition operator $T_{12}(E)$ that satisfies the 
Lippmann-Schwinger equation with the "internal"potential $v_{12}$
\begin{equation}
T_{12}(E)=  v_{12} +  v_{12} G^0(E) T_{12}(E)\;,
\end{equation}
we wright the "truncated" Green's operator in the form
\begin{equation}
G^Q_{12}(E)= G^0(E) + G^0(E) T_{12}(E) G^0(E) - P G_{12}(E)\;.
\end{equation}

The integration in the expression for the effective potential (5) is 
performed over the variables of the two-particle bound system. The 
operator $v_{eff}$ acts only on functions of variables that describe 
the relative motion of the center of mass of the two-particle bound 
complex and the particle 0. In the general case the effective potential   
$v_{eff}$ is nonlocal and dependent on the energy of the relative 
motion of the center of mass of the two-particle bound complex and 
the particle 0, $\varepsilon$.

Considering in the integral equation for the operator $R$ (6) the 
potential of the "external" interaction between the particles of 
the complex and the particle 0, which is the source of the field, $v_0$ 
as a perturbative operator, in the second order of the expansion 
of $R$ in terms of $v_0$ in (6) we obtain the formula for the 
operator of the polarization interaction potential $v_{pol}$,
\begin{equation}
v_{pol} = \langle \psi \mid v_0 G^Q_{12}(E) v_0 \mid \psi \rangle  
\end{equation}
where the "truncated" Green's operator $G^Q_{12}(E)$ is determined 
by the expression (10).

To derive the formula for the electric multipole polarizabilities of 
the complex, it is necessary to consider the behaviour of the potential (11)
at asymptotically large distances between the complex and the particle 0 
at the kinetic energy of the relative motion of the particle 0 and 
the complex $\varepsilon$ much lesser than the binding energy of 
the complex ($\varepsilon \ll b$).

According to the uncertainty principle, in the case of large (compared to 
the size of the complex) distances between the particle 0 and the center 
of mass of the complex $\rho_0$ the variable momentum of relative motion 
between the particles within the complex is much greater than the variable 
momentum of the relative motion of the particle 0 and the center of mass 
of the complex. In such a case, due to the adiabatic character of motion 
in the expressions for operators $G^0(E)$ and $G_{12}(E)$ and for the 
transition operator $T_{12}$, which are contained in the formula for the   
"truncated" Green's function $G^Q_{12}(E)$ (10), it is reasonably to neglect 
by the kinetic-energy operator $h^0_0$ as compared with the variable 
quantity descibed by the operator $h^0_{12}$, and in such a way to reduce 
the three-body problem to two-body one, 
\begin{displaymath}
G^0(E) \equiv (E - h^0_{12} - h^0_0)^{-1} \rightarrow (E - h^0_{12})^{-1}
\cdot I_0 \equiv g^0(E)\cdot I_0 \;,
\end{displaymath}
\begin{equation}
G_{12}(E) \equiv (E - h^0_{12} - h^0_0 - v_{12})^{-1}\rightarrow 
(E - h^0_{12} - v_{12})^{-1} \cdot I_0 \equiv g(E)\cdot I_0 \;,
\end{equation}
\begin{displaymath}
T_{12}(E) \rightarrow t(E) \cdot I_0 \;,
\end{displaymath}
\begin{displaymath}
G^Q_{12}(E) \equiv G_{12}(E) - P G_{12}(E) \rightarrow [g(E) - P g(E)]\cdot I_0 
\equiv g^Q(E)\cdot I_0 \;.    
\end{displaymath}
Here $I_0$ is a unit operator that acts on the functions of variables describing 
the position of the center of mass of the two-body particle relative to the 
particle 0, $g^0(E)=(E - h^0_{12})^{-1}$ and $g(E)=(E - h^0_{12} - v_{12})^{-1}$ 
are the free and complete two-particle Green's operators, and $t(E)$ is the 
two-particle transition operators. The equations for $g$ and $t$ follow from 
the three-particle equations (8) and (9) becoming as
\begin{equation}
g(E)=g^0(E)+g^0(E)v_{12} g(E)
\end{equation}
and
\begin{equation}
t(E)=v_{12}+v_{12} g^0(E) g(E)\;.
\end{equation}

Then according to (12) the formula for the three-particle Green's operator 
$G^Q_{12}$ (10) is transformed into the expression for the two-particle "truncated" 
operator
\begin{equation}
           g^Q(E) = g^0(E) + g^0(E) t(E) g^0(E) - P g(E)\;,
\end{equation}                      
and the formula for the polarization potential (11) under the adiabatic 
conditions of motion (12) and $\varepsilon = 0$ takes the form 
\begin{equation}
v_{pol} = \langle \psi \mid v_0 g^Q(-b) v_0 \mid \psi \rangle \; .
\end{equation}

Notice that the operator $g^Q(E)$ does not contain the pole singularity at $E=-b$ 
due to the mutual cancellation of the pole singularities of the operators $t(E)$ 
and $Pg(E)$ in the expression (15), which nearly the above point have the form 
\begin{equation}
      t(E) = \frac{\mid \gamma \rangle \langle \gamma \mid}{E+b} + \tilde{t}(E)\;,\;\;      
      P g(E) = \mid \psi \rangle [E - h^0_{12} - v_{12}]^{-1}\langle \psi \mid = 
      \mid \psi \rangle \frac{1}{E+b}\langle \psi \mid\;, 
\end{equation}      
where the operator $\tilde{t}$ denotes the smooth part of the transition operator 
and the function $\mid \gamma \rangle$ is the vertex function
\begin{equation}
\mid \gamma \rangle \equiv v_{12}\mid \psi \rangle = [g^0(-b)]^{-1}\mid \psi \rangle \;.
\end{equation}
Cancelling the pole  terms in (15) at $E=-b$ with the use of (17), (18) and the 
identity
\begin{equation}
 g^0(E)[g^0(-b)]^{-1}=[g^0(-b)]^{-1} g^0(E)=1-(E+b)g^0(E) 
\end{equation}
we obtain the expression for the "truncated" propagator (15) at the point $E=-b$ 
\begin{equation}
g^Q(-b)=g^0(-b)-g^0(-b)\mid \psi \rangle \langle \psi \mid -       
\mid \psi \rangle \langle \psi \mid g^0(-b) + g^0(-b)\tilde{t}(-b)g^0(-b)\;.
\end{equation}

The potential of the "external" interaction of all particles of the complex 
and the charged particle 0 that generates the electric field is
\begin{equation}
v_0= \frac{e_1 e_0}{r_{01}} +  \frac{e_2 e_0}{r_{02}} \;,
\end{equation}
where
\begin{displaymath}
{\bf r_{01}}=-\frac{m_2}{m_{12}}{\bf r}-\mbox{\boldmath $\rho$}\;,\;\;\;
{\bf r_{02}}=\frac{m_1}{m_{12}}{\bf r}-\mbox{\boldmath $\rho$}\;,\;\;\;
{\bf r=r_{12}=r_1-r_2}\;,
\end{displaymath} 
\begin{equation}
\mbox{\boldmath $\rho$}=\frac{1}{m_{12}}(m_1 {\bf r_1} + m_2 {\bf r_2}) - {\bf r_0}\;,
\end{equation}
${\bf r}$ is the radius-vector of the particle 1 relative the particle 2, 
$\mbox{\boldmath $\rho$}$ is the radius-vector of the center of mass of 
the complex relative to the charged particle 0, $m_i$ is  the mass of the 
particle $i$, $m_{12}=m_1+m_2$. At asymptotically large distances between 
the complex and the particle 0, using the expansion in power series of 
$(r/{\rho})\ll 1$ we wright the potential $v_0$ in the form
\begin{equation}
v_0= \frac{e_0}{\rho} \sum_{\lambda=0}^{\infty} (-1)^{\lambda}
\frac{M_{\lambda}({\bf r},\hat{\mbox{\boldmath $\rho$}})}{\rho^{\lambda}}\;.
\end{equation}
Here $M_{\lambda}$ is the operator of the multipole moment of the charged 
particles of the complex (multiplication operator in coordinate space)
\begin{equation}
M_{\lambda}(\mbox{\boldmath $r$},\hat{{\mbox{\boldmath $\rho$}}})= 
e_1 r^{\lambda}_{1c} P_{\lambda}(\hat{\mbox{\boldmath $r_{1c}$}} \cdot 
\hat{{\mbox{\boldmath $\rho$}}})+
e_2 r^{\lambda}_{2c} P_{\lambda}(\hat{\mbox{\boldmath $r_{2c}$}} \cdot 
\hat{{\mbox{\boldmath $\rho$}}})=
\left[e_1 \left(\frac{m_2}{m_{12}}\right)^{\lambda}+
e_2 \left(-\frac{m_2}{m_{12}}\right)^{\lambda} \right] r^{\lambda}
P_{\lambda}(\hat{\mbox{\boldmath $r$}} \cdot \hat{{\mbox{\boldmath $\rho$}}}) \;,
\end{equation}
${\bf r}_{ic}$ is the radius-vector of the charged particle of the 
complex relative to its center of mass ($i=1,2$), 
$$
{\bf r}_{1c}=\frac{m_2}{m_{12}}{\bf r}, \qquad {\bf r}_{2c}=-\frac{m_1}{m_{12}}{\bf r}\;, 
$$
$P_{\lambda}(x)$ is the Legendre polinomial, and the unit vectors are 
marked with a hat, $\hat{\bf a}={\bf a}/a$.

Substituting the asymptotical expression for $v_0$ (23) in Eq. (16) we 
write the polarization potential as 
\begin{equation}
v_{pol}(\rho) = \left( \frac{e_0}{\rho}\right)^2 \sum_{\lambda,{\lambda}^{\prime} = 0}^{\infty} 
\left(-\frac{1}{\rho}\right)^{\lambda + {\lambda}^{\prime}} 
\langle \psi \mid M_{\lambda} g^Q(-b) M_{{\lambda}^{\prime}}\mid \psi \rangle \;,
\end{equation}
where the matrix element is determined according to Eq. (20) by the formula
\begin{equation}
\begin{array}{rcl}
\langle \psi\mid M_{\lambda} g^Q(-b) M_{{\lambda}^{\prime}}\mid \psi \rangle & = &
\langle\psi\mid M_{\lambda} g^0(-b) M_{{\lambda}^{\prime}}\mid \psi \rangle \\[3mm] \nonumber
& - & \langle \psi\mid M_{\lambda} g^0(-b) \mid \psi \rangle 
\langle \psi\mid M_{{\lambda}^{\prime}}\mid \psi \rangle  \\[3mm] \nonumber
& - & \langle \psi\mid M_{\lambda}\mid \psi \rangle 
\langle \psi\mid g^0(-b) M_{{\lambda}^{\prime}}\mid \psi \rangle \\[3mm]
& + & \langle \psi\mid M_{\lambda} g^0(-b) \tilde{t}(-b) 
g^0(-b)M_{{\lambda}^{\prime}} \mid \psi \rangle \;.\nonumber
\end{array}
\end{equation}

The general formula (26) that determines the polarization interaction potential 
for the quantum two-particle complex is considerably simplified when applying it 
to specific (atomic and nuclear) systems. If, for example, the potential of 
interaction between the particles of the complex $v_{12}(r)$ is invariant 
relative to the space reflection ${\bf r} \rightarrow -{\bf r}$, then
the wave function of the bound complex is characterized by a definite parity.
The conservation of the parity leads to nullification of the matrix elements 
\begin{equation}
\langle\psi\mid M_{\lambda}\mid \psi\rangle \;,\quad  \langle\psi\mid M_{\lambda} 
g^0(-b) \mid \psi\rangle \;\quad \mbox{ and } \;\quad \langle\psi\mid g^0(-b) 
M_{\lambda}\mid \psi\rangle 
\end{equation}
(for example, for the systems with the Coulomb or nuclear interactions). 
In this case the wave function of the bound complex is characterized by 
a definite parity. The conservation of the parity leads to nullification 
of the matrix elements which are present in the second and third summands 
of the formula (26), at odd values of $\lambda$.

The formula (26) is also simplified if the interaction potential $v_{12}$ is 
invariant relative to spatial rotations, that is central. In such a case 
the wave function of the complex $\psi$ is characterized by a certain 
value of the orbital angular momenta $l$, and the matrix elements (27) 
also vanish, if the triangle condition $\Delta(l\lambda l)$ does not obey. 
Specifically, they vanish after integrating in angular variables for all 
$\lambda\neq 0$, if the total orbital moment of the complex is equal 
to zero, $l=0$. 
\\ [.2in]
{\samepage
\noindent {\bf 3. Electric multipole polarizabilities of the hydrogen-like atom}\\
\nopagebreak

Let us apply the approach developed in the above to hydrogen-like atoms
(atomic systems consisting of one electron (the particle 1) 
and a nucleus (the particle 2)). The interaction between the particles is 
described by the Coulomb potential  
\begin{equation}
v_{12}(r)=v^C(r)= \frac{e_1 e_2}{r}\;,
\end{equation}
where $e_1=-e$ is the charge of the electron, $e_2=Ze$ is the charge of 
the nucleus ($e$ is the elementary positive charge), $Z$ is the atomic 
number of the nucleus. }

We describe the atomic system in the momentum space. The radius-vector 
operator in this case is ${\bf r}= i\hbar \nabla_{\bf p}=i\nabla_{\bf k}$, 
${\bf k}$ being the wave vector. The Coulomb interaction potential
\begin{equation}
\langle {\bf k} \mid v^C \mid {\bf k^{\prime}}\rangle
 =\int d{\bf r} v^C(r) \exp \left[ -i({\bf k}-{\bf k^{\prime}}){\bf r}\right]  
=-\frac{4\pi Ze^2}{\mid {\bf k}-{\bf k^{\prime}}\mid ^2} 
\end{equation}
expanded in the spherical harmonics of the variable wave vectors 
${\bf k}\mbox{ and }{\bf k^{\prime}}$ corresponding to the orbital angular 
momentum $l$ and magnetic $m$ quantum numbers has the form 
\begin{equation}
\langle {\bf k}\mid v^C \mid {\bf k^{\prime}}\rangle = 4\pi\sum_{l=0}^{\infty} 
\sum_{m=-l}^{l} v^C_l(k,k^{\prime}) Y_{lm}(\hat{\bf k}) 
Y^{*}_{lm}(\hat{\bf k}^{\prime})\;.
\end{equation}

The wave function of the $S$-wave bound state of the atom   
\begin{equation}
\psi (k)= 4\pi \int_{0}^{\infty} dr r^2 j_0(kr) \psi(r)
\end{equation}
satisfies the integral Schr\"{o}dinger equation
\begin{equation}
\psi = g^{0}(-b)v^C_0 \mid \psi \;,
\end{equation}
where $g^{0}(-b)=(-b-h^0)^{-1}$ is the free Green's operator, $b$ is the 
binding energy of the atom, $v^C_0$ is the operator 
of the $S$-wave partial component of the Coulomb interaction potential 
in the expansion (30),
\begin{equation}
 b=\frac{{\hbar}^2{\kappa}^2}{2\mu_{12}}\;, \qquad \kappa=\frac{Z\mu_{12}e^2}{{\hbar}^2}\;,
\end{equation}
$\mu_{12}=m_1 m_2 /m_{12}$ is the reduced mass of the electron and 
the nucleus ($m_1 = m$ is the mass of the electron, $m_2$ is the 
mass of the nucleus).

In the following, we shall restrict our consideration in the 
hydrogen-like atom with the nucleus of the infinitely great 
mass as compared to the mass of the electron 
($m_1 /m_2 \rightarrow 0, \; \mu_{12} \rightarrow 0$), neglecting
the second term in the expression (24).

In the coordinate space the action of the multipole moment operator 
$M_{\lambda}$ on the the bound-state wave function of the atom is 
reduced to the multiplication 
\begin{equation}
\langle {\bf r}| M_{\lambda}|\psi \rangle =  
M_{\lambda}({\bf r},\hat{{\mbox{\boldmath$\rho$}}})\psi(r), \qquad 
M_{\lambda}({\bf r},\hat{{\mbox{\boldmath$\rho$}}})=e_1 r^{\lambda}
P_{\lambda}(\hat{{\bf r}} \cdot \hat{{\mbox{\boldmath $\rho$}}}) \;.
\end{equation}
In the momentum space the formula (34) takes the form
\begin{equation}
\langle {\bf k} \mid M_{\lambda} \mid \psi \rangle = 
(-i)^{\lambda} e_1 \varphi_{\lambda}(k) 
P_{\lambda}(\hat{{\bf k}} \cdot \hat{{\mbox{\boldmath $\rho$}}}) \;,
\end{equation}
where
\begin{equation}
\varphi_{\lambda}(k)= 4\pi \int_{0}^{\infty} dr r^{\lambda+2} j_{\lambda}(kr) \psi(r)\;.
\end{equation}

Applying the Rayleigh formula [27] for the spherical Bessel functions
\begin{equation}
j_{\lambda}(kr) = (-1)^{\lambda} \frac{k^{\lambda}}{r^{\lambda}} 
\left( \frac{1}{k} \frac{d}{dk} \right)^{\lambda} j_0(kr)\;,  
\end{equation}
the function ${\varphi}_{\lambda}(k)$ can be written as
\begin{equation}
{\varphi}_{\lambda}(k) = (-1)^{\lambda} k^{\lambda}
\left( \frac{1}{k} \frac{d}{dk} \right)^{\lambda} \psi (k)\;.
\end{equation}
From the formula (38), it follows the recurrence expression that
relates the functions ${\varphi}_{\lambda +1}$ and ${\varphi}_{\lambda}$:
\begin{equation}
{\varphi}_{\lambda+1}(k) = \frac{\lambda}{k}{\varphi}_{\lambda}(k)- 
\frac{{d \varphi}_{\lambda}(k)}{dk} \quad
\mbox{ proceeding from }\quad {\varphi}_0(k)= \psi(k)\;.
\end{equation}

Using Eqs. (25), (26) and (35) we can determine the polarization potential 
of the interaction of the hydrogen-like atom and the particle 0. In this case 
the first term in the right-hand side of Eq. (26) is given by 
\begin{equation}
<\psi\mid M_{\lambda} g^0(-b) M_{{\lambda}^{\prime}}\mid \psi> =
-\frac{e^2}{2\lambda+1} \int_{0}^{\infty}\frac{k^2 dk}{2\pi^2}
\frac{\mid \varphi_{\lambda}(k)\mid ^2}{\frac{k^2}{2m}+b} 
{\delta}_{\lambda {\lambda}^{\prime}}\;.
\end{equation}
The matrix elements in the second and third terms in Eq. (26) 
are written as
\begin{equation}
\langle \psi \mid M_{\lambda} \mid \psi \rangle = -e \delta_{\lambda 0}\;,
\end{equation}
\begin{equation}
\langle \psi\mid  g^0(-b)M_{\lambda} \mid \psi \rangle 
= \langle \psi \mid M_{\lambda} g^0(-b) \mid \psi \rangle 
= e \frac{R_1}{b} {\delta}_{\lambda 0} \;, 
\end{equation}  
where
\begin{equation}
R_1 = b \int_{0}^{\infty}\frac{k^2 dk}{2\pi^2}
\frac{\mid \psi(k)\mid ^2}{\frac{k^2}{2m}+b} =
b \int_{0}^{\infty}\frac{k^2 dk}{2\pi^2}
\frac{<\gamma\mid k><k\mid \gamma>}{\left(\frac{k^2}{2m}+b\right)^3}  
\end{equation}
and $<k\mid \gamma>$ is the vertex function (18). 

Note that the smooth part of the Coulomb transition operator (defined 
according to (17)) that is contained in Eq. (26) at the 
energy of the bound state of the complex, $\tilde{t}(-b)\equiv \tilde{t}^C(-b)$, 
consists of two operators --- a part of the $S$-wave partial component (that corresponds to 
the orbital momenta $l=0$), $\tilde{t}^C_0(-b)$, and the sum of all higher partial orbital 
components (with $l\geq 1$) at the same energy, $t^C_h(-b)$,
\begin{equation}
\tilde{t}^C(-b) = \tilde{t}^C_0(-b) + t^C_h(-b) \;.  
\end{equation} 
As shown in the preceding work [21], the operator $\tilde{t}^C_0(E)$ at 
the point $E=-b$ has the factorable form
\begin{equation}
\tilde{t}^C_0(-b) = \mid \gamma \rangle \left( -\frac{R_1}{b} \right)
\langle \gamma \mid \;.      
\end{equation}   
The matrix elements of the operator $t^C_h(-b)$ can be written in the form 
of the expansion over the set of the spherical functions of the angles of 
the momentum variables
\begin{equation}
<{\bf k}|t^C_h(-b)|{\bf k^{\prime}}>=4\pi\sum_{l=1}^{\infty} \sum_{m=-l}^{l}
t^C_l(k,k^{\prime}; -b) Y_{lm}(\hat{\bf k})Y^{*}_{lm}(\hat{\bf k}^{\prime})\;.
\end{equation}

Taking into consideration the relations (35), (41) and (44) -- (46), we find 
the fourth term in Eq. (26) in the form of the sum of the matrix element 
\begin{equation}
\langle \psi \mid M_{\lambda} g^0(-b) \tilde{t}^C_0(-b) g^0(-b) 
M_{{\lambda}^{\prime}} \mid \psi \rangle = 
e^2 \left( - \frac{R_1}{b} \right) {\delta}_{\lambda 0} 
{\delta}_{\lambda^{\prime} 0}
\end{equation}
and the matrix element
\begin{displaymath}
\langle \psi \mid M_{\lambda} g^0(-b) t^C_h(-b) g^0(-b) 
M_{{\lambda}^{\prime}} \mid \psi \rangle 
\end{displaymath}
\begin{equation}
= \left( 1 - \delta_{\lambda 0} \right) \left( 1 - \delta_{\lambda^{\prime} 0} \right) 
\delta_{\lambda \lambda^{\prime}} \frac{e^2}{2\lambda + 1}
\int_{0}^{\infty} \frac{k^2 dk}{2\pi^2} \int_{0}^{\infty} \frac{{k^{\prime}}^2 dk^{\prime}}{2\pi^2} 
\frac {\varphi^{*}_{\lambda}(k)\varphi_{\lambda}(k^{\prime})}{\left( \frac{k^2}{2m}+b \right) 
\left( \frac{{k^{\prime}}^2}{2m}+b \right)} t^C_{\lambda}(k,k^{\prime}; -b)\;.
\end{equation}

Substituting the expressions (40) -- (42), (47) and (48) into Eq. (26) and 
taking into account that the matrix elements (47) with $\lambda=0, 
\lambda^{\prime}=1$ and $\lambda=1, \lambda^{\prime}=0$ equal to zero 
and the matrix elements with $\lambda = \lambda^{\prime}=0$ are mutually cancelled,
we wright the polarization potential of interaction between the particle 
0 and the atom (25) in the form
\begin{equation}
v_{pol}(\rho) = - \frac{e_0^2}{2} \sum_{\lambda = 1}^{\infty} 
\frac{\alpha_{E\lambda}}{\rho^{2\lambda + 2}}\;,
\end{equation}
where the coefficients
\begin{equation}
\alpha_{E\lambda} = -2 \langle \psi \mid M_{\lambda} g^Q(-b) M_{\lambda} 
\mid \psi \rangle\;,  
\end{equation}
which represent the electric $2^{\lambda}-$pole polarizabilities of 
the atom are determined by the expression 
\begin{displaymath}
\alpha_{E\lambda} = \frac{2}{(2\lambda+1)} e^2 
\left\{ \int_{0}^{\infty}\frac{dk k^2}{2\pi^2} \frac{\mid \varphi_{\lambda}(k)
\mid ^2}{\frac{\hbar^2 k^2}{2 m}+b} \right. \\ [-2mm] \nonumber
\end{displaymath} 
\begin{equation}
 - \left. \int_{0}^{\infty}\frac{dk k^2}{2\pi^2} 
\int_{0}^{\infty} \frac{dk^{\prime}k^{{\prime}2}}{2\pi^2} 
\frac{\varphi^{*}_{\lambda}(k) t^C_{\lambda}(k,k^{\prime};-b)
\varphi_{\lambda}(k^{\prime})}{\left( \frac{\hbar^2 k^2}{2 m}+b \right) 
\left( \frac {\hbar^2 {k^{\prime}}^2}{2 m}+b \right)} \right\}\;. 
\end{equation}

The formula for the electric $2^{\lambda}$-pole polarizability of the 
hydrogen-like atom (51) contains the function $\varphi_{\lambda}(k)$ 
that, according to (38), is expressed in terms of the derivatives 
(up to the order $\lambda$, inclusive) of the wave function of the 
bound state of the atom with respect to the relative momentum 
variable $k$. For example, the corresponding functions for the dipole 
($\lambda=1$), quadrupole ($\lambda=2$) and octupole ($\lambda=3$) 
polarizabilities  have the form
\begin{equation}
\begin{array}{rcl}
\varphi_{1}(k)& = & - \psi^{\prime}(k) \;, \\ 
\varphi_{2}(k)& = & - \frac{1}{k} \psi^{\prime}(k) + \psi^{\prime\prime}(k)  \;, \\
\varphi_{3}(k)& = & - \frac{3}{k^2} \psi^{\prime}(k) + \frac{3}{k} \psi^{\prime\prime}(k) 
- \psi^{\prime\prime\prime}(k) \;.
\end{array}
\end{equation} 

The second term of the expression for the multipole polarizability 
$\alpha_{E\lambda}$ (51) describes the contribution from the virtual 
scattering of the particles of the system in intermediate states. 
It contains the partial component of the Coulomb transition matrix 
at the negative energy of the bound state of the atom 
$t^C_{\lambda}(k,k^{\prime};-b)$ that satisfies the Lippmann-Schwinger 
integral equation 
\begin{equation}
t^C_{\lambda}(k,k^{\prime};-b) = v^C_{\lambda}(k,k^{\prime})-
\int_{0}^{\infty} \frac{dk^{\prime\prime} {k^{\prime\prime}}^2}{2\pi^2}
v^C_{\lambda}(k,k^{\prime\prime}) \frac{1}{\frac{{k^{\prime\prime}}^2}{2m}+b}
t^C_{\lambda}(k^{\prime\prime},k^{\prime};-b)\;\;, \\[3mm]  
\end{equation}     
where $v^C_{\lambda}(k,k^{\prime})$ is the partial component of the Coulomb 
interaction potential. 

For the Coulomb transition matrix with the negative energy 
$t_{\lambda}^C(k^,k^{\prime};-b)$, some representations have been derived --- 
in the form of a sum [37], of an integral [38], with explicitly separated 
singularities in the variables of the transfer momentum and the energy [22] 
and others. They can be applied for determining $\alpha_{E\lambda}$ by the 
formula (51) numerically. In particular, that has been performed so in our 
preceding paper [21] for the dipole, quadrupole and octupole polarizabilities 
of the hydrogen atom using the corresponding partial components of the  
Coulomb transition matrix derived in Ref. [22]. 

In this paper, leaning upon the t-matrix formalism, we are working out 
a new method for rigorous analytical determination of $\alpha_{E\lambda}$ 
for the hydrogen-like atoms. 

Applying the integral transformations with the kernels
\begin{equation}
{\cal K}^0_{\lambda}(k,k^{\prime};-b)=-\frac{1}{2{\pi}^2} 
\frac{{k^{\prime}}^2}{\frac{{k^{\prime}}^2}{2m}+b}
v^C_{\lambda}(k,k^{\prime}) \mbox{   and   } 
{\cal K}_{\lambda}(k,k^{\prime};-b)=-\frac{1}{2{\pi}^2} 
\frac{{k^{\prime}}^2}{\frac{{k^{\prime}}^2}{2m}+b}
t^C_{\lambda}(k,k^{\prime};-b)
\end{equation}
that transform the function $\varphi_{\lambda}(k)$, which describes 
$2^{\lambda}$-pole distortion of the system, into the functions
\begin{equation}
f_{\lambda}(k) = - \int_{0}^{\infty} \frac{dk^{\prime} {k^{\prime}}^2}{2\pi^2}
v^C_{\lambda}(k, k^{\prime}) \frac{1}{\frac{{k^{\prime}}^2}{2 m}+b} \varphi_\lambda(k^{\prime}) 
\end{equation}
and
\begin{equation}
\phi_{\lambda}(k) = - \int_{0}^{\infty} \frac{dk^{\prime} {k^{\prime}}^2}{2\pi^2}
t^C_{\lambda}(k,k^{\prime};-b) \frac{1}{\frac{{k^{\prime}}^2}{2 m}+b} \varphi_\lambda(k^{\prime}) \;,
\end{equation}
we derive from the integral equation for t-matrix (53) the inhomogeneous 
integral equation for the function $\phi_{\lambda}(k)$ in the form
\begin{equation}
\phi_{\lambda}(k) = f_{\lambda}(k) - \int_{0}^{\infty} \frac{dk^{\prime} {k^{\prime}}^2}{2\pi^2}
v^C_{\lambda}(k,k^{\prime})\frac{1}{\frac{{k^{\prime}}^2}{2 m}+b} \phi_{\lambda}(k^{\prime}) 
\end{equation}
in which the free term is the function $f_{\lambda}(k)$ and the kernel is the same as the kernel  
of the integral equation for $t^C$-matrix (53), ${\cal K}^0_{\lambda}(k,k^{\prime};-b)$. 
In such a case, the formula (51) for the electric $2^{\lambda}$-pole polarizability of 
the hydrogeh-like atom takes the form 
\begin{equation}
\alpha_{E\lambda} = \frac{2}{(2\lambda+1)\pi^2} \frac{me^2}{\hbar^2}
\int_{0}^{\infty} dk k^2 \frac{\varphi_{\lambda}^{*}(k) \left[ \varphi_{\lambda}(k) 
+ \phi_{\lambda}(k) \right]}{k^2 + \kappa^2} \;.
\end{equation}

Note, that the function $\phi_{\lambda}(k)$ in Eq. (58) describes the 
contribution into the $2^{\lambda}$-pole polarizability of the bound complex 
from multiple scattering of its constituents in the intermediate state with the 
orbital momentum $\lambda$ and the negative energy $-b$ (scattering completely off 
the energy shell). This function will be referred to as the off-shell scattering 
function. Below we show that the integral equation (57) determining the function 
$\phi_{\lambda}(k)$ can be solved by the rigorous analytical way. 
\\ [.2in]
{\samepage
\noindent {\bf 4. Analytical solving of the integral equation for the 
function $\phi_{\lambda}(k)$ \\ 
in the case of the ground bound state}\\
\nopagebreak

In the momentum space, in accordance with Eqs. (31) and (33), the normalized wave 
function of the hydrogen-like atom in the ground bound state ($n=1,\;\; l=0$) 
is given by 
\begin{equation}
\psi^{10}(k) = \frac{8\sqrt{\pi}\kappa^{5/2}}{(k^2 + \kappa^2)^2}\;, \qquad
\kappa = \frac{Zme^2}{\hbar^2}\;. 
\end{equation} }

The partial component of the expansion of the Coulomb interaction potential (29)
into the spherical functions of the orbital angular momentum (30), which is contained 
in the kernel of the main integral equation (57), is qiven by the expression 
\begin{equation}
v^C_{\lambda}(k,k^{\prime}) = \frac{2\pi e_1 e_2}{k k^{\prime}} Q_{\lambda} 
\left( \frac{k^2+{k^{\prime}}^2}{2 k k^{\prime}} \right)\;,
\end{equation}
where the function $Q_{\lambda}(x)$ is the Legendre function of the second kind [28] 
\begin{equation}
Q_{\lambda}(x) = \frac{1}{2} P_{\lambda}(x) \ln \left( \frac{x+1}{x-1}\right)
 - W_{\lambda-1}(x)\;,
\end{equation}
\begin{displaymath}
W_{-1}(x)=0\;, \qquad  W_{\lambda-1}(x)=\sum_{k=1}^{\lambda} \frac{1}{k} P_{\lambda-k}(x) P_{k-1}(x)  
\end{displaymath}

Substituting the expression for the wave function (59) into the formula (38) 
or (39), we obtain the function $\varphi_{\lambda}^{10}(k)$ in the form 
\begin{equation}
\varphi_{\lambda}^{10}(k) = 2^{\lambda+3} (\lambda+1)! \sqrt{\pi} {\kappa}^{5/2}
\frac{k^{\lambda}}{(k^2+{\kappa}^2)^{\lambda+2}}\;.
\end{equation}

Using the expressions for the partial component of the interaction potential (60) 
and for the function $\varphi_{\lambda}^{10}$ (62), we find the free term of the 
integral equation (57) in the form  
\begin{equation}
f_{\lambda}^{10}(k) = 2^{\lambda+4} (\lambda+1)! \frac{1}{\sqrt{\pi}}{\kappa}^{7/2} 
\frac{1}{k} {\cal F}_{\lambda}^{10}(k)\;,
\end{equation}
where 
\begin{displaymath}
{\cal F}_{\lambda}^{10}(k) \equiv \int_{0}^{\infty} dk^{\prime}
\frac{{k^{\prime}}^{\lambda + 1}}{({k^{\prime}}^2+{\kappa}^2)^{\lambda+3}} 
Q_{\lambda} \left( \frac{k^2+{k^{\prime}}^2}{2 k k^{\prime}} \right) 
\end{displaymath}
\begin{equation}
= \frac{\pi}{4(\lambda+1)(\lambda+2){\kappa}^3}
\frac{k^{\lambda+1}\left[ k^2 + (2\lambda+3){\kappa}^2 \right]} 
{(k^2+{\kappa}^2)^{\lambda+2}}\;.
\end{equation} 

Substituting the integration result (64) into Eq. (63), we obtain the explicit 
expression for the function $f_{\lambda}^{10}(k)$ 
\begin{equation}
f_{\lambda}^{10}(k) = 2^{\lambda+2} \frac{\lambda!}{\lambda+2} \sqrt{\pi} {\kappa}^{1/2}
\frac{k^{\lambda}\left[ k^2 + (2\lambda+3){\kappa}^2 \right]}{(k^2+{\kappa}^2)^{\lambda+2}}\;.
\end{equation}

Taking into account Eq. (60), we write the integral equation for the function 
$\phi_{\lambda}^{10}(k)$ (57) as
\begin{equation}
\phi_{\lambda}^{10}(k) = f_{\lambda}^{10}(k) +\frac{2\kappa}{\pi} \frac{1}{k} 
\int_{0}^{\infty} dk^{\prime} \frac{1}{{k^{\prime}}^2+{\kappa}^2}Q_{\lambda} 
\left( \frac{k^2+{k^{\prime}}^2}{2 k k^{\prime}} \right) \phi_{\lambda}^{10}(k^{\prime}) 
\end{equation}

In view of the construction of the free term $f_{\lambda}^{10}(k)$ (63) that is 
determined by the integral ${\cal F}_{\lambda}^{10}(k)$ (64), we seek for the solution 
of the integral equation (66) in the form of an expression that contains a specific 
two-term factor 
\begin{equation}
\phi_{\lambda}^{10}(k) = C_{\lambda} \frac{k^{\lambda}}{(k^2+{\kappa}^2)^{\lambda+2}}
\left[ A_{\lambda} k^2 + B_{\lambda} {\kappa}^2 \right]\;, 
\end{equation}
where the coefficient of the expression for the free term (65) was separated out 
explicitly,
\begin{equation}
C_{\lambda}= 2^{\lambda+2} \frac{\lambda!}{\lambda+2} {\sqrt{\pi}} {\kappa}^{1/2}\;,
\end{equation}
while the coefficients $ A_{\lambda}$ and $ B_{\lambda}$ should be determined. 

Substituting the function (67) into the equation (66), we obtain the equality
\begin{equation}
\frac{k^{\lambda}}{(k^2+{\kappa}^2)^{\lambda+2}}
\left[ A_{\lambda} k^2 + B_{\lambda} {\kappa}^2 \right] = 
\frac{k^{\lambda}\left[ k^2+(2\lambda + 1){\kappa}^2 \right] }
{(k^2+{\kappa}^2)^{\lambda+2}} + \frac{2\kappa}{\pi} \frac{1}{k} 
\left[ A_{\lambda} {\cal E}_{\lambda}^{10}(k) + B_{\lambda}{\kappa}^2 
{\cal F}_{\lambda}^{10}(k)\right] \;
\end{equation}
where the integral ${\cal E}_{\lambda}^{10}(k)$ is distinct from the integral 
${\cal F}_{\lambda}^{10}(k)$ (64) only by the factor $k^{\prime 2}$ in the 
integrand, 
\begin{displaymath}
{\cal E}_{\lambda}^{10}(k) \equiv \int_{0}^{\infty} dk^{\prime}
\frac{{k^{\prime}}^{\lambda + 3}}{({k^{\prime}}^2+{\kappa}^2)^{\lambda+3}} 
Q_{\lambda} \left( \frac{k^2+{k^{\prime}}^2}{2 k k^{\prime}} \right) 
\end{displaymath}
\begin{equation}
= \frac{1}{4(\lambda+1)(\lambda+2) \kappa} 
\frac{k^{\lambda+1}\left[ (2\lambda+3)k^2 + {\kappa}^2 \right]} 
{(k^2+{\kappa}^2)^{\lambda+2}}\;.
\end{equation}
Taking the values of the integrals (64) and (70) and carring out required 
simplifications, we wright the relation (69) as
\begin{displaymath}
 A_{\lambda} k^2 + B_{\lambda} {\kappa}^2  = 
\left[ k^2+(2\lambda + 1){\kappa}^2 \right] + \frac{1}{2(\lambda+1)(\lambda+2)}
\left\{ A_{\lambda} \left[ (2\lambda+3)k^2 + {\kappa}^2 \right]  \right.
\end{displaymath}
\begin{equation}
\left. + B_{\lambda} \left[ k^2 + (2\lambda+3){\kappa}^2 \right] \right\}\;.
\end{equation}
Applying the relationship (71) at two values of the variable $k$, say,
at the points $k=0$ and $k\rightarrow \infty$, we find hence the set
of two coupled linear algebraic equations for the coefficients $A_{\lambda}$ and 
$B_{\lambda}$:
\pagebreak
\begin{displaymath}
(2\lambda^2 + 4\lambda +1) A_{\lambda} - B_{\lambda} = 2(\lambda+1)(\lambda+2) \;
\end{displaymath}
\begin{equation}
A_{\lambda} - (2\lambda^2 + 4\lambda +1) B_{\lambda}
= - 2(\lambda+1)(\lambda+2)(2\lambda + 3) 
\end{equation}
the solution of which is
\begin{equation}
A_{\lambda} = \frac{\lambda + 2}{\lambda}\;, \qquad  
B_{\lambda} = \frac{(2\lambda+1)(\lambda+2)}{\lambda}\;.
\end{equation}
It is worth noting that alternative equations arising with the use of other 
values of the variable $k$ in the relation (71) present linear combinations 
of the equations (72).

Inserting the obtained values (73) into Eq. (67), we find the solution of 
the integral equation (66) in the form
\begin{equation}
\phi_{\lambda}^{10}(k) = 2^{\lambda+2} (\lambda-1)! \sqrt{\pi}{\kappa}^{1/2} 
\frac{k^{\lambda}}{(k^2+{\kappa}^2)^{\lambda+2}} 
\left( k^2+(2\lambda + 1){\kappa}^2 \right) \;. 
\end{equation} 
\\ [.2in]
{\samepage
\noindent {\bf 5. Electric multipole polarizabilities of the hydrogen-like atom \\
in the ground state}\\
\nopagebreak

Substituting the expressions (62) and (74) for the functions 
$\varphi_{\lambda}^{10}(k)$ and $\phi_{\lambda}^{10}(k)$ into the formula (58), 
we obtain the final result for the electric $2^{\lambda}$-pole polarizabilities 
of the hydrogen-like atom in the ground state 
\begin{equation}
{\alpha}_{E\lambda}(10) = \frac{(\lambda + 2)(2\lambda+1)!}  
{2^{2\lambda} \lambda} \frac{1}{{\kappa}^{2\lambda+1}}\;.
\end{equation} }

The formula (75) is coincident with the result derived using the perturbative 
Dalgarno-Lewis technique [9] which reduces the problem of taking into consideration 
of the contributions from all excited intermediate states to solving a corresponding 
inhomogeneous differential equation that, as it was shown in Ref. [9], in 
the case of the Coulomb interaction may be performed analytically.

To analyze contributions of intermediate virtual states into the polarizabilities 
$\alpha_{E\lambda}$ leaning upon our general formula (51), let us separate out the 
Born term from the partial Coulomb t-matrix 
\begin{equation}
t^C_{\lambda}(k,k^{\prime};-b) = v^C_{\lambda}(k,k^{\prime}) + 
\Delta t^C_{\lambda}(k,k^{\prime};-b) 
\end{equation}
and taking into consideration Eqs. (55) and (56) write the polarizability 
$\alpha_{E\lambda}$ as a sum of three terms,
\begin{equation}
{\alpha}_{E\lambda} = {\alpha}_{\lambda}^0 + {\alpha}_{\lambda}^{BS} + 
{\alpha}_{\lambda}^{MS}\; ,
\end{equation}
where the first term ${\alpha}_{\lambda}^0$ contains the free virtual propagator 
\begin{equation}
\alpha_{\lambda}^0 = \frac{2\kappa}{(2\lambda+1)\pi^2} 
\int_{0}^{\infty} dk k^2 \frac{ \mid \varphi_{\lambda}(k)\mid ^2}{k^2 + {\kappa}^2}\;,
\end{equation}
the term ${\alpha}_{\lambda}^{BS}$ describes the Born (single) scattering in the intermediate 
state, and the term ${\alpha}_{\lambda}^{MS}$ characterizes the rest of contributions 
from the multiple scattering (of the two-fold order and higher),
\pagebreak
\begin{displaymath}
\alpha_{\lambda}^{BS} = \frac{2\kappa}{(2\lambda+1)\pi^2} 
\int_{0}^{\infty} dk k^2 \frac{ \varphi_{\lambda}(k) f_{\lambda}(k)}{k^2 + {\kappa}^2} \;,
\end{displaymath}
\begin{equation}
\alpha_{\lambda}^{MS} = \frac{2\kappa}{(2\lambda+1)\pi^2} 
\int_{0}^{\infty} dk k^2 \frac{ \varphi_{\lambda}(k) 
\left[ \phi_{\lambda}(k) - f_{\lambda}(k) \right]}{k^2 + {\kappa}^2} \;.
\end{equation}

Making use of (62), (65) and (74), we find for the ground state
\begin{displaymath}
\alpha_{\lambda}^0(10) = \frac{(\lambda+1)(2\lambda+5)(2\lambda)!}{2^{2\lambda}(\lambda+2) 
{\kappa}^{2\lambda+1}}\;, \qquad
\alpha_{\lambda}^{BS}(10) = 
\frac{(\lambda+3)(2\lambda+3)!}{2^{2\lambda + 1}(\lambda+1)(\lambda+2)^2 
(2\lambda+1){\kappa}^{2\lambda+1}}\;, 
\end{displaymath}
\begin{equation}
\alpha_{\lambda}^{MS}(10) =
\frac{(2{\lambda}^2+9\lambda+8)(2\lambda)!}{2^{2\lambda} \lambda 
(\lambda+2)^2 {\kappa}^{2\lambda+1}}\;.
\end{equation}

As an illustration, we show in Table 1 the values of the components 
${\alpha}_{\lambda}^0(10)$, ${\alpha}_{\lambda}^{BS}(10)$ and 
${\alpha}_{\lambda}^{MS}(10)\;$ together with the corresponding 
electric $2^{\lambda}$-pole polarizabilities of the hydrogen-like atom 
in the ground state ${\alpha}_{E\lambda}(10)$ for $\lambda$ = 1,2,3 and 4.

\vspace*{.2in} 
{\samepage
{\normalsize \tablename\hspace{2mm}1.\hspace{1mm} The components 
$\alpha_{\lambda}^0(10)$, $\alpha_{\lambda}^{BS}(10)$ and $\alpha_{\lambda}^{MS}(10)$ 
determining the electric dipole ($\lambda=1$), quadrupole ($\lambda=2$), 
octupole ($\lambda=3$) and hexadecapole ($\lambda=4$) polarizabilities 
of the hydrogen-like atom in the ground state, $\alpha_{E\lambda}(\mbox{10})$, calculated with 
the use of the direct t-matrix approach  (in ${\kappa}^{-(2\lambda+1)}$)
\nopagebreak  
\begin{center} \begin{tabular}{|c|c|c|c|c|} \hline
\multicolumn{1}{|c|}{}& \multicolumn{1}{c|}{}& \multicolumn{1}{c|}{}&
\multicolumn{1}{c|}{}&
\multicolumn{1}{c|}{} \\
\multicolumn{1}{|c|}{$\lambda$}&
\multicolumn{1}{c|}{$\alpha_{\lambda}^0(10)$}&
\multicolumn{1}{c|}{$\alpha_{\lambda}^{BS}(10)$}&
\multicolumn{1}{c|}{$\alpha_{\lambda}^{MS}(10)$}&
\multicolumn{1}{c|}{$\alpha_{E\lambda}(\mbox{10})$} \\
\multicolumn{1}{|c|}{}&
\multicolumn{1}{c|}{}&
\multicolumn{1}{c|}{}&
\multicolumn{1}{c|}{}&
\multicolumn{1}{c|}{} \\ \hline
\multicolumn{1}{|c|}{}&
\multicolumn{1}{c|}{}&
\multicolumn{1}{c|}{}&
\multicolumn{1}{c|}{}&
\multicolumn{1}{c|}{} \\
$1$&$\frac{7}{3}$&$\frac{10}{9}$&$\frac{19}{18}$&$\frac{9}{2}$ \\
\multicolumn{1}{|c|}{}&
\multicolumn{1}{c|}{}&
\multicolumn{1}{c|}{}&
\multicolumn{1}{c|}{}&
\multicolumn{1}{c|}{} \\
$2$&$\frac{81}{8}$&$\frac{105}{32}$&$\frac{51}{32}$&$15$\\
\multicolumn{1}{|c|}{}&
\multicolumn{1}{c|}{}&
\multicolumn{1}{c|}{}&
\multicolumn{1}{c|}{}&
\multicolumn{1}{c|}{} \\
$3$&$99$&$\frac{243}{10}$&$\frac{159}{20}$&$\frac{525}{4}$ \\ 
\multicolumn{1}{|c|}{}&
\multicolumn{1}{c|}{}&
\multicolumn{1}{c|}{}&
\multicolumn{1}{c|}{}&
\multicolumn{1}{c|}{} \\
$4$&$\frac{6825}{4}$&$\frac{2695}{8}$&$\frac{665}{8}$&$\frac{8505}{4}$ \\
\multicolumn{1}{|c|}{}&
\multicolumn{1}{c|}{}&
\multicolumn{1}{c|}{}&
\multicolumn{1}{c|}{}&
\multicolumn{1}{c|}{} \\  \hline

\end{tabular}
\end{center}}
\bigskip }

The data of Table 1 indicate that the dipole polarizability 
of the hydrogen-like atom $\alpha_{E\lambda}(10)$ in the ground state 
consists of two almost equal parts --- of the first term containing 
the free propagator, $\alpha_1^0(10) = \frac{7}{3}$, and of the sum of two 
other terms that describe single and multiple virtual scattering 
in the $P$-wave orbital state, $\alpha_1^{BS}(10) + \alpha_1^{MS}(10) = \frac{13}{6}$. 
In the case of the polarizabilities of the higher polarity the 
contribution of the term with the free propagator still escalates 
coming up to 67.5 \% 
, 75.4 \% 
and 80.2 \% 
respectively for the quadrupole, octupole and hexadecapole polarizabilities. 
The contribution to the polarizability $\alpha_{E\lambda}(10)$ from 
the Born term, $\alpha_1^{BS}(10)$, decreases with $\lambda$ from 24.7 \% 
for $\lambda = 1$ to 15.8 \% 
for  $\lambda = 4$. The contribution of the term that describes 
the multiple virtual scattering, $\alpha_{\lambda}^{MS}(10)$,  decreases with 
$\lambda$ more distinctly: from the value 23.5 \% 
for $\lambda = 1$ to 3.9 \% 
for $\lambda = 4$.

It follows that the contribution into the polarizability $\alpha_{E\lambda}(10)$ 
(77) from the first two terms --- with the free propagator $\alpha_{\lambda}^0(10)$ 
and describing the single virtual scattering $\alpha_{\lambda}^{BS}(10)$ --- is 
dominant taking the values 76.5 \% 
, 89.4 \%  
and 93.9 \%  
in the case of the dipole ($\lambda = 1$), quadrupole ($\lambda = 2$) 
and octupole ($\lambda = 3$) polarizabilities, respectively. 
\\ [.2in]
{\samepage
\noindent {\bf 6. Electric dipole polarizability of the hydrogen-like atom\\
in the excited $2S$-state}\\ 
\nopagebreak

The normalized wave function of the hydrogen-like atom in the excited bound 
state with $n=2, l=0$ is of the form
\begin{equation}
\psi^{20}(k) = \sqrt{8\pi}\kappa^{5/2} 
\frac{\left.k^2 - \frac{1}{4} \kappa^2 \right.}{\left( k^2 + \frac{1}{4} \kappa^2 \right) ^3}\;.
\end{equation} }

The electric dipole polarizability of the hydrogen-like atom 
in the excited $2S$-state is determined by the expression (58) with $\lambda=1$, 
the function $\varphi_1 = \varphi_1^{20}$,
\begin{equation}
\varphi_1^{20}(k) =  - \frac{d\hspace{1ex}}{dk}\psi^{20}(k) =  
4\sqrt{8\pi}\kappa^{5/2} \frac{k \left( k^2 - \frac{1}{2}\kappa^2 \right)}
{\left( k^2 + \frac{1}{4} \kappa^2\right) ^4}
\end{equation}  
and the off-shell scattering function $\phi_1(k)=\phi_1^{20}(k)$. The last-mentioned 
function satisfies the integral equation (57) the free term of which $f_1^{20}(k)$ is 
determined by the formula (55),
\begin{displaymath}
f_1^{20}(k)= \int_{0}^{\infty} dk^{\prime} {\cal K}^0_1(k,k^{\prime};-b_{20}) 
\varphi_1^{20}(k^{\prime}) 
\end{displaymath}
\begin{equation}
= C_1^{20} \frac{k\left[ \left( k^2 -\frac{3}{4} \kappa^2 \right) ^2 
+ \frac{7}{2} \kappa^4 \right]}{\left( k^2 + \frac{1}{4} \kappa^2 \right) ^4} \;,
\qquad C_1^{20} = - \frac{\sqrt{8\pi}}{3} {\kappa}^{1/2}\;.
\end{equation}  

The solution of the integral equation for the function $\phi_1^{20}(k)$, 
\begin{equation}
\phi_1^{20}(k) = f_1^{20}(k) + \frac{2\kappa}{\pi k}  
\int_{0}^{\infty} dk^{\prime} \frac{1}{{k^{\prime}}^2+ \frac{1}{4}{\kappa}^2}Q_1 
\left( \frac{k^2+{k^{\prime}}^2}{2 k k^{\prime}} \right) \phi_1^{20}(k^{\prime})\;, 
\end{equation}
we seek in the form of a binomial constructed from 
$\left( k^2 -\frac{3}{4} \kappa^2 \right) ^2$ and $\kappa^4$,
\begin{equation}
\phi_1^{20}(k)= \frac{C_1^{20}k}{\left( k^2 + \frac{1}{4} \kappa^2 \right) ^4 }
\left[ \bar{A}_1 \left( k^2 -\frac{3}{4} \kappa^2 \right) ^2 + \bar{B}_1 \kappa^4 \right],
\end{equation}  
similarly as it was done in the case of the ground state when the integral 
equation (66) was analytically solved leaning upon the expression (67).

Substituting the function (85) into the equation (84) we obtain 
\pagebreak   
\begin{displaymath}
\left[ \bar{A}_1 \left( k^2 -\frac{3}{4} \kappa^2 \right) ^2 + \bar{B}_1 \kappa^4 \right] 
= \left[ \left( k^2 -\frac{3}{4} \kappa^2 \right) ^2 + \frac{7}{2} \kappa^4 \right] 
\end{displaymath}
\begin{equation}
+ \frac{2\kappa}{\pi k^2} \left\{ \bar{A}_1 {\cal G}_1^{20}(k) 
- \frac{3}{2} \bar{A}_1 \kappa^2 {\cal E}_1^{20}(k) + \left( \frac{9}{16} \bar{A}_1 
+ \bar{B}_1 \right){\kappa}^4 {\cal F}_1^{20}(k) \right\} 
\left( k^2 + \frac{1}{4} \kappa^2 \right)^4 \;, 
\end{equation}  
where the following designations for arising integrals are used 
\begin{displaymath}
{\cal G}_1^{20}(k) \equiv \int_{0}^{\infty} dk^{\prime}
\frac{{k^{\prime}}^6}{({k^{\prime}}^2+\frac{1}{4}{\kappa}^2)^5} 
Q_1 \left( \frac{k^2+{k^{\prime}}^2}{2 k k^{\prime}} \right) \;
\end{displaymath}
\begin{displaymath}
= \; \frac{\pi}{96\kappa}\frac{k^2}{\left( k^2 + \frac{1}{4}{\kappa}^2 \right)^4} 
\left[ 35 k^4 + \frac{7}{2}{\kappa}^2 k^2 + \frac{3}{16}{\kappa}^4 \right]\;, 
\end{displaymath}
\begin{displaymath}
{\cal E}_1^{20}(k) \equiv \int_{0}^{\infty} dk^{\prime}
\frac{{k^{\prime}}^4}{({k^{\prime}}^2+\frac{1}{4}{\kappa}^2)^5} 
Q_1 \left( \frac{k^2+{k^{\prime}}^2}{2 k k^{\prime}} \right) \;
\end{displaymath}
\begin{equation}
= \; \frac{\pi}{24 \kappa^3}\frac{k^2}{\left( k^2 + \frac{1}{4}{\kappa}^2 \right)^4} 
\left[ 5 k^4 + \frac{17}{2}{\kappa}^2 k^2 + \frac{5}{16}{\kappa}^4 \right]\;, 
\end{equation}
\begin{displaymath}
{\cal F}_1^{20}(k) \equiv \int_{0}^{\infty} dk^{\prime}
\frac{{k^{\prime}}^2}{({k^{\prime}}^2+\frac{1}{4}{\kappa}^2)^5} 
Q_1 \left( \frac{k^2+{k^{\prime}}^2}{2 k k^{\prime}} \right) \;
\end{displaymath}
\begin{displaymath}
= \; \frac{\pi}{6 \kappa^5}\frac{k^2}{\left( k^2 + \frac{1}{4}{\kappa}^2 \right)^4} 
\left[ 3 k^4 + \frac{7}{2}{\kappa}^2 k^2 + \frac{35}{16}{\kappa}^4 \right]\;.
\end{displaymath} 

Using the relations (86) and (87) at the points $k=0$ and $k\rightarrow \infty$, 
we deduce the set of two coupled algebraic equations for the coefficients 
$\bar{A}_1$ and $\bar{B}_1$ --- in analogy to the derivation of the 
set (72) for the atom in the ground state:
\begin{equation}
\begin{array}{rcl}
3 \bar{A}_1 & + & \frac{13}{3} \bar{B}_1 = 65\;,\\[.1in]
\frac{1}{3} \bar{A}_1 & - & \bar{B}_1 = 1 \;.
\end{array}
\end{equation}
The solution of the set (88) is
\begin{equation}
\bar{A}_1 = \frac{78}{5}\;, \qquad \bar{B}_1 = \frac{21}{5}\;.
\end{equation} 
In this way, the solution of the integral equation for the off-shell 
scattering function $\phi_1^{20}(k)$ (84) takes the form
\begin{equation}
\phi_1^{20}(k)= \frac{3}{5} C_1^{20} 
\frac{k\left[ 26 \left( k^2 -\frac{3}{4} \kappa^2 \right) ^2 
+ 7 \kappa^4 \right]}{\left( k^2 + \frac{1}{4} \kappa^2 \right) ^4} \;,
\qquad C_1^{20} = - \frac{\sqrt{8\pi}}{3} \kappa^{1/2}\;.
\end{equation}  

Substituting the expressions (82) and (90) for the functions 
$\varphi_{1}^{20}(k)$ and $\phi_1^{20}(k)$ into the formula (58) with 
$\lambda=1$, we deduce the value of the electric dipole polarizability 
of the hydrogen-like atom in the excited $2S$-state (with $n=2, \; l=0$) 
\begin{equation}
\alpha_{E1}(20) = \frac{2712}{5} \frac{1}{\kappa^3}\;.
\end{equation}  

Analyzing the role of the scattering in the intermediate virtual state 
and separating out the Born term  from the $P$-wave partial Coulomb 
transition matrix according to Eq. (76) we write the electric dipole 
polarizability of the hydrogen-like atom in the excited $2S$-state 
in the form 
\begin{equation}
\alpha_{E1}(20) = \alpha_1^0(20) +  \alpha_1^{BS}(20) + \alpha_1^{MS}(20)\;,
\end{equation}
The values of individual components in Eq. (92) we derive using 
the formulas (78) and (79) and the obtained above functions 
$\varphi_1^{20}(k)$, $f_1^{20}(k)$ and $\phi_1^{20}(k)$ (Eqs. 
(82), (83) and (90)):
\begin{equation}
\alpha_1^0(20) = \frac{502}{3}\; \frac{1}{\kappa^3}\;, \qquad
\alpha_1^{BS}(20) = \frac{1205}{9}\; \frac{1}{\kappa^3}\;, \qquad
\alpha_1^{MS}(20) = \frac{10853}{45}\; \frac{1}{\kappa^3}\;.
\end{equation}

For comparison we show in Table 2 the values of the components
$\alpha_1^0(n0)$, $\alpha_1^{BS}(n0)$ and $\alpha_1^{MS}(n0)$ 
and the electric dipole polarizability of the hydrogen-like 
atom in the ground ($n=1$, $l=0$) and excited ($n=2$, $l=0$) 
$S$-wave states. 

\vspace*{.1in} 
{\samepage
{\normalsize \tablename\hspace{2mm}2.\hspace{1mm} The components 
$\alpha_1^0(n0)$, $\alpha_1^{BS}(n0)$ and  $\alpha_1^{MS}(n0)$ 
determining the electric dipole polarizabilities 
of the hydrogen-like atom $\alpha_{E1}(nl)$ in the ground ($n=1$, $l=0$) 
and excited ($n=2$, $l=0$) states, derived with the use of the direct 
t-matrix approach (in ${\kappa}^{-3}$) 
\nopagebreak
\begin{center} \begin{tabular}{|c|c|c|c|c|} \hline
\multicolumn{1}{|c|}{}& \multicolumn{1}{c|}{}& \multicolumn{1}{c|}{}&
\multicolumn{1}{c|}{}&
\multicolumn{1}{c|}{} \\
\multicolumn{1}{|c|}{$n$}&
\multicolumn{1}{c|}{$\alpha_1^0(n0)$}&
\multicolumn{1}{c|}{$\alpha_1^{BS}(n0)$}&
\multicolumn{1}{c|}{$\alpha_1^{MS}(n0)$}&
\multicolumn{1}{c|}{$\alpha_{E1}(n0)$} \\
\multicolumn{1}{|c|}{}&
\multicolumn{1}{c|}{}&
\multicolumn{1}{c|}{}&
\multicolumn{1}{c|}{}&
\multicolumn{1}{c|}{} \\ \hline
\multicolumn{1}{|c|}{}&
\multicolumn{1}{c|}{}&
\multicolumn{1}{c|}{}&
\multicolumn{1}{c|}{}&
\multicolumn{1}{c|}{} \\
$1$&$\frac{7}{3}$&$\frac{10}{9}$&$\frac{19}{18}$&$\frac{9}{2}$ \\
\multicolumn{1}{|c|}{}&
\multicolumn{1}{c|}{}&
\multicolumn{1}{c|}{}&
\multicolumn{1}{c|}{}&
\multicolumn{1}{c|}{} \\
$2$&$\frac{502}{3}$&$\frac{1205}{9}$&$\frac{10853}{45}$&$\frac{2712}{5}$\\
\multicolumn{1}{|c|}{}&
\multicolumn{1}{c|}{}&
\multicolumn{1}{c|}{}&
\multicolumn{1}{c|}{}&
\multicolumn{1}{c|}{} \\  \hline

\end{tabular}
\end{center}}
\bigskip }

Note that for the electric dipole polarizabilities of the atom 
in the excited state $\alpha_{E1}(20)$ the contribution of  
the terms which describe the Born and multiple scattering in 
virtual states, $\alpha_1^{BS}+\alpha_1^{MS}$, exceeds that of  
the term with the free virtual propagator, $\alpha_1^0$, as distinct 
from the polarizabilities of the atom in the ground state   
$\alpha_{E1}(10)$ for which the mentioned contributions are almost 
equal. 
\\ [.2in]
{\samepage
\noindent {\bf 7. Conclusions and outlook}\\
\nopagebreak

To summarize, the t-matrix approach to description of the polarization interaction 
of the quantum two-particle bound system (both atomic and nuclear) that arises under influence of 
the external electric field has been developed in this work. We show that
in the case of the Coulomb system the application of the t-matrix description 
permits to perform the rigorous analytical solution of the corresponding 
Lippmann-Schwinger integral equation that determines the electric 
multipole polarizabilities of the hydrogen-like atom in the ground state. 
The obtained formula for the electric polarizability in this case agrees 
with the Dalgarno-Lewis result [9], which has been derived in the framework 
of the modified Rayleigh-Schr\"{o}dinger perturbation theory by solving 
analytically an inhomogeneous differential equation that describes the 
correction to the unperturbated state for the contributions from all the 
intermediate bound and continuum virtual states. }

The most important advantage of the t-matrix formalism for determination of the 
electric polarizability consists in the fact that its application allows to avoid 
cumbersome calculation of contributions from a great many discrete and continuum 
excited virtual states. Instead, it is necessary to determine one or a few 
(depending on the form of the interaction) partial t-matrix components.
An essential circumstance in this case is also that the t-matrix elements are 
contained only at the negative energy of the bound state and present therefore 
real functions of two variable momenta. 

In the case of the simplest composite nucleus --- the deuteron, when the 
$S$-wave partial interaction between the constituents of the complex 
is dominated and the contribution from the $P$-wave interaction is only
slight, the electric polarizability is determined mainly by the free 
propagator. The corresponding calculations of the electric dipole, 
quadrupole and octupole polarizabilities of the deuteron have recently been 
performed with the application of the t-matrix formalism in Ref. [18]. 

The t-matrix method applied in this paper can be directly generalized to 
systems with more complicated interactions between the constituents. The 
presence of the tensor interaction can cause the anisotropic polarization 
properties of the system. For example, the electric dipole or quadrupole 
polarizabilities of the deuteron nucleus will be differed in the cases 
when the external electric field is directed parallel or transversely to 
the spin of the deuteron [31]. 
 
It is also promising to apply the t-matrix approach for analytical 
determination of the electric polarizabilities of the hydrogen-like atoms 
in excited (degenerated) states. 

Another important direction of investigation on the polarization effects 
is the calculation of the van der Waals interactions between atoms using 
the t-matrix formalism. It is known [32], the direct precision measurement 
of the van der Waals interaction between two neutral isolated atoms in which 
one of electrons is in a highly excited state with large principal quantum 
number $n$ (in the 50--100 range)--- so called Rydberg atoms. The obtained 
new data on fine details of the van der Waals interaction require in-depth 
theoretical interpretation [33].

Finally, that is of the particular importance to study the possibility of 
the generalization of the proposed approach for more complex three- and 
four-particle quantum systems, which are described the Faddeev and 
Faddeev-Yakubovsky integral equations [34, 35]. 

In nuclear physics, the electric polarizabilities of the few-body nuclei 
are important characteristics that contain additional independent 
information on the fundamental nuclear force. Unfortunately, existing 
value for the the electric dipole polarizability of the nucleus
$^3$He extracted from the data on the cross-sections for photodisintegration 
of the nucleus [36, 37] ($\alpha_{E1}=0.15\pm0.02$ fm$^3$)and obtained by way 
of the experimental study of deviations from Rutherford scattering in elastic 
scattering of $^3$He nuclei from $^{208}$Pb at energies well below the 
Coulomb barrier [38] ($\alpha_{E1}=0.25\pm0.04$ fm$^3$) are not to be in 
agreement among themselves. The application of the transition-matrix approach to 
the description of the deformation properties of the three-nucleon nuclei 
$^3$H and $^3$He in the electric field on the basis of the Faddeev 
mathematical formalism will help to determine reliably the values of the 
dipole polarizabilities of these nuclei. \\ [.2in]

\vspace*{.1in} 
\noindent {\footnotesize {\bf References} 
\vspace*{.1in}
\begin{itemize} 
\setlength{\baselineskip}{.1in} 
\item[{\tt [1]}]{\sc W. Pauli}, {\it Z. Phys.}, {\bf 36} (1926) 336-363. 
\item[{\tt [2]}]{\sc E. Schr\"{o}dinger}, {\it Ann. Phys. Leipzig}, {\bf 80} (1926) 437-490. 
\item[{\tt [3]}]{\sc P. S. Epstein}, {\it Phys. Rev.}, {\ 28} (1926) 695-710. 
\item[{\tt [4]}]{\sc G. Wentzel}, {\it Z. Phys.}, {\bf 38} (1926) 518-529. 
\item[{\tt [5]}]{\sc I. Waller}, {\it Z. Phys.}, {\bf 38} (1926) 635-646. 
\item[{\tt [6]}]{\sc J. W. S. Rayleigh}, "Theory of Sound" 2nd Edition, Vol. 1,  
                p.p. 115-118, Macmillan, London, 1894. 
\item[{\tt [7]}]{\sc H. A. Bethe and E. S. Salpeter}, in "Handbuch der Physik," 
                Vol. 35, Section {\bf 25}$\beta$, Springer, Berlin, 1957. 
\item[{\tt [8]}]{\sc L. Castillejo, I. C. Percival and M. J. Seaton}, 
                {\it Proc. Roy. Soc. A} {\bf 254}, No. 1277 (1960) 259-272. 
\item[{\tt [9]}]{\sc A. Dalgarno and J. T. Lewis}, {\it Proc. Roy. Soc. A} {\bf 233}, 
                No. 1192 (1955) 70-74. 
\item[{\tt [10]}]{\sc C. Schwartz}, {\it Ann. Phys. NY} {\bf 6}(1959) 156-169. 
\item[{\tt [11]}]{\sc L. I. Schiff}, "Quantum Mechanics" 3nd Edition, p.p. 266-267, 
                McGraw-Hill, New York, 1968. 
\item[{\tt [12]}]{\sc M. A. Maize, M. A. Antonacci and F. Marsiglio}, {\it 
                Am. J. Phys.} {\bf 79}(2011) 222-225. 
\item[{\tt [13]}]{\sc V. A. Fock}, {\it Z. Phys.}, {\bf 98} (1935) 145-154.            
\item[{\tt [14]}]{\sc H. A. Becker and K. Bleuler}, {\it Z. Naturforschung}, 
                {\bf 31a} (1976) 517-523. 
\item[{\tt [15]}]{\sc V. F. Kharchenko, S. A. Shadchin and S. A. Permyakov}, 
                {\it Physics Letters B}, {\bf 199} (1987) 1-4. 
\item[{\tt [16]}]{\sc V. F. Kharchenko and S. A. Shadchin}, "Three-body Theory of 
                the Effective Interaction between a Particle and a Two-Particle 
                Bound System," Preprint ITP-93-24E, Institute for Theoretical 
                Physics, Kyiv, 1993. 
\item[{\tt [17]}]{\sc V. F. Kharchenko and S. A. Shadchin}, {\it Ukrainian Journal 
                of Physics}, {\bf  42} (1997) 912-920. 
\item[{\tt [18]}]{\sc V. F. Kharchenko}, {\it Int. J. Mod. Phys. E}, {\bf 22} 
                (2013) 1350031, 1-13; nucl-th/1209.2004.
\item[{\tt [19]}]{\sc V. F. Kharchenko and A. V. Kharchenko}, {\it Collected Physical 
                Papers Lviv}, {\bf 7} (2008) 432-443; nucl-th/0811.2565.             
\item[{\tt [20]}]{\sc V. F. Kharchenko and A. V. Kharchenko}, {\it Int. J. Mod. 
                Phys. E}, {\bf 19} (2010) 225-242; nucl-th/1003.5769. 
\item[{\tt [21]}]{\sc V. F. Kharchenko}, {\it J. Mod. Phys.}, {\bf 4} (2013) 99-107; 
                nucl-th/1208.1394.
\item[{\tt [22]}]{\sc S. A. Shadchin and V. F. Kharchenko}, {\it J. Phys. B}, 
                {\bf 16} (1983) 1319-1322. 
\item[{\tt [23]}]{\sc N. C. Francis and K. M. Watson}, {\it Phys. Rev.}, {\bf 92} 
                (1953) 291-303. 
\item[{\tt [24]}]{\sc H. Feshbach}, {\it Ann. Phys. NY} {\bf 5}(1958) 357-390. 
\item[{\tt [25]}]{\sc H. Feshbach}, {\it Ann. Phys. NY} {\bf 19}(1962) 287-313. 
\item[{\tt [26]}]{\sc C. J. Joachain}, "Quantum Collision Theory," North-Holland,  
                Amsterdam, 1975. 
\item[{\tt [27]}]{\sc M. Abramowitz and I. A. Stegun}, "Handbook of Mathematical 
                Functions," Dover, NY, 1965. 
\item[{\tt [28]}]{\sc V. F. Bratsev and E. D. Trifonov}, {\it Vest. Leningrad. 
                Gos. Univ.} {\bf 16} (1962) 36-39. 
\item[{\tt [29]}]{\sc J. Schwinger}, {\it J. Math. Phys.} {\bf 5}(1964) 1606-1608.
\item[{\tt [30]}]{\sc I. S. Gradshtein and I. M. Ryzhik}, "Tables of Integrals, Sums, 
                Series and Products," Fizmatgiz, Moscow, 1962. 
\item[{\tt [31]}]{\sc A. V. Kharchenko}, {\it Nuclear Physics A} {\bf 617}(1997) 34-44. 
\item[{\tt [32]}]{\sc L. Beguin, A. Vernier, R. Chicireanu, T. Lahaye and A. Browaeys}, 
                {\it Phys. Rev. Lett.}, {\bf 110}(2013) 263201, 1-5. 
\item[{\tt [33]}]{\sc M. Weidem\"{u}ller}, {\it Physics} {\bf 6}(2013) 71-75. 
\item[{\tt [34]}]{\sc L. D. Faddeev}, {\it Soviet Physics JETP} {\bf 12} (1961) 1014-1019. 
\item[{\tt [35]}]{\sc O. A. Yakubovsky}, {\it Soviet Journal of Nuclear Physics} 
                {\bf 5} (1967) 937-942. 
\item[{\tt [36]}]{\sc G. A. Rinker}, {\it Phys. Rev. A} {\bf 14} (1976) 18-29. 
\item[{\tt [37]}]{\sc V. D. Efros, W.Leidemann and G. Orlandini}, 
                {\it Phys. Lett. B} {\bf 408} (1997) 1-6; nucl-th/9707007. 
\item[{\tt [38]}]{\sc F. Goeckner, L. O. Lamm and L. D. Knutson}, 
                {\it Phys. Rev. C} {\bf 43} (1991) 66-72.                 

\end{itemize}} 

\end{document}